# WIGGLER FOR CESR OPERATION AT 2GEV[2]

A.A. Mikhailichenko

*Wilson Laboratory, Cornell University, Ithaca, NY 14853*

For low energy operation strategy we advocate utilization of many short wigglers in contrast with single long wiggler. This allows begin to operate very naturally with few strong field wigglers giving necessary damping time on expense of energy spread. By adding more and more wigglers in the ring, as these wigglers are manufactured and tuned, the field in the wigglers will be decreased, keeping necessary damping. This strategy allows the mostly effective operation of CESR with minimum down time. This also gives flexibility in operation in wider energy scale without non-reversible modifications.

## SCALING IN SR

The formulas represented below are given only for the purposes of scaling law definition. In real model of machine all these parameters are calculated with numerical codes.

Natural energy spread in the beam as a function of energy could be estimated as the following. The rate of energy losses in magnets is the following

$$\frac{dE}{dt} \cong \frac{2}{3} r_0 mc^3 \frac{\gamma^4}{\rho^2}, \tag{1}$$

$r_0 = e^2/mc^2$, $\gamma = E/mc^2$, $\rho$ is the bending radius. The last gives the damping time as

$$\tau_s \cong \frac{E}{dE/dt} = \frac{3}{2} \frac{\rho^2}{r_0 c \gamma^3} \sim \frac{1}{\gamma^3}.$$

So the number of turns required for damping goes to

$$n_s \cong \frac{c\tau_s}{2\pi\rho} = \frac{3}{4\pi} \frac{\rho}{r_0} \frac{1}{\gamma^3}.$$

As each particle radiates $\sim \alpha = e^2/\hbar c$ photons with characteristic energy $\hbar\omega \cong \frac{3}{2}\frac{\hbar c}{\rho}\gamma^3$ at formation length $\cong \rho/\gamma$, the total number of photons radiated per damping time goes to

$$N_\gamma \cong \alpha \frac{2\pi\rho}{\rho/\gamma} n_s = 2\pi\alpha\gamma \cdot n_s = \frac{3}{2}\frac{\alpha\rho}{r_0\gamma^2} = \frac{3}{2}\frac{\rho}{\lambdabar_C \gamma^2} \sim \frac{1}{\gamma^2},$$

where $\lambdabar_C = r_0/\alpha = \hbar/mc = 3.86\cdot 10^{-13} m$ is Compton wavelength. The energy spread in the beam goes respectively to

$$\frac{\Delta E}{E} \cong \frac{\sqrt{N_\gamma} \cdot \hbar\omega}{E} = \sqrt{\frac{3\rho}{2\lambdabar_C}} \frac{3\hbar\gamma}{2mc\rho} = \left(\tfrac{3}{2}\right)^{3/2} \gamma \sqrt{\frac{\lambdabar_C}{\rho}} \sim \gamma.$$

---

[1] This paper in its original version was distributed in December 2000 as internal report and placed at web page under CESR-C upgrades. January 22, 2001 the final version was placed at the same web page. Recently as this web page was reorganized, I decided to put this paper as CBN report for more availability. This program paper has historical interest however. In this paper the modular design for CESR-c wigglers originated. Free space inventory for support of modular design was made earlier in November 2000. In this paper all key elements of wiggler calculations and design considered, such as optimization of shapes for the coils and poles. Cold mass design was introduced also. In this paper odd/even pole schemes for wiggler analyzed too. Taking into account specifics of pretzel operation, the conclusion was made in favor of odd pole-wiggler. To minimize nonlinearities, tapering in the pole field strength was recommended as ½, 1.

[2] Electronic version available at http://www.lns.cornell.edu/public/CBN/2004/CBN04-7/CESR_2GeV.pdf

The last formula gives the scaling law. As the energy spread in CESR at 5GeV is ~ 0.0005, defined mostly by hard bend magnets, corresponding energy spread in the beam at 2 *GeV* will be $(\Delta E/E)_{2GeV} \cong 2 \cdot 10^{-4}$. This energy spread is small, as it does not include the energy spread generated by wigglers.

The last could be estimated as the following. First, let us substitute into formula (1) the bending radius in a wiggler field expressed through the wiggler parameter $K = eH_w \lambdabar_w / mc^2 \cong 93.4 \cdot H_w[T] \cdot 2\pi\lambdabar_w[m]$, where $2\pi\lambdabar_w$ is the wiggler period as the following

$$\rho = \lambdabar_w \gamma / K.$$

So the number of turns for damping goes to

$$n_s \cong \frac{3}{2} \frac{\lambdabar_w^2}{r_0 \gamma K^2 L_w},$$

where $L_w$ is the length of the wiggler, and we suggested that only damping device is a wiggler. Length of formation goes to $l_f \cong \rho/\gamma \cong \lambdabar_w / K$ and the number of photons per pass goes to

$$N_{\gamma,pass} \cong \alpha \frac{L_w}{l_f} \cong \alpha K \frac{L_w}{\lambdabar_w}.$$

To obtain total number of photons one needs to multiply the last number by beam population. Energy of quanta goes to $\hbar\omega \cong \tfrac{3}{2}\hbar c K \gamma^2 / \lambdabar_w$, so the energy spread goes to

$$\frac{\Delta E}{E} \cong \sqrt{N_{\gamma,pass} \cdot n_s} \frac{\hbar\omega}{E} = \sqrt{\frac{\alpha K L_w}{\lambdabar_w} \frac{3\lambdabar_w^2}{2 r_0 \gamma K^2 L_w}} \cdot \frac{3\hbar c K \gamma^2}{2\lambdabar_w mc^2 \gamma} = \left(\tfrac{3}{8}\right)^{1/2} \sqrt{K \frac{\lambdabar_C \gamma}{\lambdabar_w}}.$$

As we would like to keep the damping time constant, $n_s = const$.

$$\frac{\Delta E}{E} \cong \sqrt{N_{\gamma,pass} \cdot n_s} \frac{\hbar\omega}{E} = \sqrt{n_s \frac{\alpha K L_w}{\lambdabar_w} \cdot \frac{3\lambdabar_C K \gamma}{2\lambdabar_w}} \propto \sqrt{H_w}$$

The bunch length could be expressed as the following

$$\sigma_b \cong R_0 \eta \frac{\omega_0}{\Omega_s} \frac{\Delta E}{E},$$

where $R_0$ is average CESR radius, $|\eta| = \frac{E}{R_0} \frac{\partial R_0}{\partial E}$, $\Omega_s \cong \omega_0 \cdot \sqrt{\frac{qeV\eta\sin\varphi_s}{2\pi E}}$ is synchrotron frequency, $\omega_0$ is revolution frequency, $q$ is RF harmonics number. So, for fixed energy

$$\sigma_b \propto \sqrt{\frac{H_w}{qV}},$$

and any increase of magnetic field could be compensated by effective voltage increase~$qV$. As the losses of energy for 2 *GeV* mode is less, than for 5*GeV*, the existing power of 500 *MHz* RF is enough for the losses compensation.

Shortening of the bunch can be made by additional superconducting cavities, operating at 1.5 *GHz* (or even 3*GHz*). Here each Mega-Volt is equivalent to 3 (or 6) *MeV* @ 500 *MHz*. This RF can be phased only for shortening of the bunch, not for compensation of SR.

These high frequency cavities are also preferable due to lack of free space, required for 500 *MHz* cavities.

## LENGTHENING IN A WIGGLER

Dispersion in the wiggler field could be represented as

$$D = \frac{K \lambdabar_w}{\gamma}.$$

The lengthening of trajectory *by one wiggler period* could be represented as

$$\Delta = \int_0^\lambda \sqrt{1 + \frac{D^2}{\lambdabar_w^2} Cos^2 \frac{s}{\lambdabar_w}} ds - \lambda_w \cong \int_0^\lambda (1 + \frac{D^2}{2\lambdabar_w^2} Cos^2 \frac{s}{\lambdabar_w}) ds - \lambda = \frac{\pi D^2}{2 \lambdabar_w} = \frac{\pi K^2 \lambdabar_w}{2\gamma^2}.$$

For parameters $\lambda = 28cm$, *B*=3*T*, (so *K*=78.5), $\gamma = 5000$, $D \cong \frac{78.5 \cdot 4.46}{5000} = 0.07 cm$ and, hence, $\Delta \cong \frac{\pi D^2}{2\lambdabar} \cong \frac{\pi 49 \cdot 10^{-4}}{2 \cdot 4.46} \cong 1.7 \cdot 10^{-3} cm$ For full single 7-pole wiggler unit it goes to $\Delta_w \cong 5 \cdot 10^{-3} cm$. Ten wigglers with ten meters total give 0.05*cm*~0.5*mm*. While increasing the number of wigglers, the magnetic field drops $\cong 1/\sqrt{L_w}$ keeping damping time constant. So the lengthening per period will drop proportionally to the length, keeping the total lengthening constant however, as the total number of wigglers increased proportionally.

## EMITTANCE GROWTH DUE TO THE SCATTERING ON RESIDUAL GAS

For the multiple scattering we have the formula

$$<\vartheta^2> \cong \left(\frac{13.6}{pc[MeV]}\right)^2 \frac{t}{X_0},$$

where the thickness of scattering medium *t* is expressed in radiation lengths. For Air at atmospheric pressure, $t/X_0 = 1$ corresponds to ~340 meters. At pressure *P*[Torr], the last distance goes to $l[m] = 340 \times 760 / P[Torr]$. So, the equilibrium emittance gain goes to

$$\Delta\varepsilon[cm \cdot rad] \cong \beta[cm] \left(\frac{13.6}{pc[MeV]}\right)^2 \frac{(c\tau_s)[m]}{340} \times \frac{P[Torr]}{760},$$

where $\beta$ is envelope function, *c* is speed of light and $\tau_s$ is a damping time. For 2 *GeV*, $\beta \sim 10m = 1000 cm$, $\tau_s = 30 msec$ and $P \cong 10^{-8} Torr$, the last formula gives

$$\Delta\varepsilon \cong 1000 \left(\frac{13.6}{2000}\right)^2 \frac{3 \times 10^8 \cdot 30 \times 10^{-3}}{340} \times \frac{10^{-8}}{760} \cong 1.7 \times 10^{-8} cm \cdot rad.$$

This number might be higher, if the ions are accumulated at the orbit. Still, this effect has an academic interest mostly in our case.

## FOCUSING IN A WIGGLER

Wiggler does not focus in radial direction due to $\int Bds = 0$. The vertical tune shift can be estimated as usual

$$\Delta Q_y \cong -\frac{1}{4\pi}\int \beta_y(s)\Delta K_y ds,$$

where $\Delta K_y \cong \frac{e\Delta G}{pc} \cong \frac{\Delta G}{(HR)}$, $(HR)$ is a magnetic rigidity, $\Delta G$ is a variation in gradient due to the pole's fringe fields. As the particle's trajectory has the slope $x' \cong K/\gamma$, the effective gradient acting to the particle over the edge is

$$\int \Delta Gds \cong \frac{KH_\perp}{\gamma} \equiv \frac{KeH_\perp \lambdabar_w mc^2}{e\lambdabar_w mc^2 \gamma} = \frac{mc^2 K^2}{e\lambdabar_w \gamma} = \frac{(HR)K^2}{\lambdabar_w \gamma^2}.$$

So effective tune shift per edge goes to

$$\Delta Q \cong -\frac{1}{4\pi}\int \beta(s)\Delta K_x ds \cong -\frac{\beta_0(s)}{4\pi(HR)}\int \Delta Gds = \frac{\beta_0(s)}{4\pi(HR)} \frac{(HR)K^2}{\lambdabar_w \gamma^2} = \frac{\beta_0(s)K^2}{4\pi\lambdabar_w \gamma^2}.$$

For $N$-pole wiggler this number goes to $\Delta Q_N \cong (N-1)\cdot \Delta Q$, so

$$\Delta Q_{wiggler} \cong \frac{L_w}{\lambda_w} \frac{\beta_0(s)K^2}{4\pi \lambdabar_w \gamma^2} \cong \frac{L_w}{\lambda_w} \frac{\beta_0(s)K^2}{2\lambdabar_w \gamma^2} \cong \beta_0(s)\frac{L_w K^2}{8\pi^2 \lambdabar_w^2 \gamma^2}.$$

As we keep the number of turns for damping $n_s \cong \frac{3}{2}\frac{\lambdabar_w^2}{r_0 \gamma K^2 L_w}$ =const,

$$\Delta Q_{wiggler} \cong \frac{\beta_0(s)}{2}\frac{L_w K^2}{\lambdabar_w^2 \gamma^2} \cong \frac{\beta_0(s)}{3r_0\gamma^3}\cdot \frac{2r_0 L_w K^2 \gamma}{3\lambdabar_w^2} = \frac{\beta_0(s)}{3r_0\gamma^3}n_s.$$

For parameters $\lambda_w = 20 cm$, B=2T, (so K=37.6), $\gamma = 5000$ (2.5 GeV), N=7, $\beta_0 \cong 10m = 1000cm$

$$\Delta Q_{wiggler} \cong \frac{\beta_0(s)}{2}\frac{L_w K^2}{\lambdabar_w^2 \gamma^2} \cong \frac{10^3 \cdot 100 \cdot 1.4 \cdot 10^3}{2\cdot 10.1 \cdot 25 \cdot 10^6} \cong 0.27 /\text{wiggler}.$$

This value can be easily compensated by neighboring lens, as

$$\int \Delta Gds \cong (N-1)\frac{(HR)K^2}{\lambdabar_w \gamma^2} = 6\frac{8\cdot 10^3 (37.6)^2}{3.18\cdot 25\cdot 10^6} = 0.85 kG.$$

As the quad's effective length is $L_{eff} \sim 60cm$, the gradient variation required goes to

$$\Delta G_{quad} \cong \frac{\int \Delta Gds}{L_{eff}} \cong \frac{0.85}{60} \cong 0.014 kG/cm,$$

while typical focusing gradient is 0.5 $kG/cm$. This value will go down together with the number of wigglers installed So the focusing is not a problem here.

Sectoring of poles also possible, as the tapering angle here is $x' \cong K/\gamma \cong 16 mrad$. The sectoring will eliminate both vertical and radial focusing. The wiggler design becomes a bit complicated however.

## LIFETIME, EMITTANCE AND ENERGY SPREAD

Parameters of the ring calculated with numerical codes, which take into account IBS and Toushek lifetime for real optics.
Emittance dynamics defined by usual equations (averaging over period)

$$\frac{d\varepsilon_x}{dt} \cong \left\langle \left( H_x + \frac{\beta_x}{\gamma^2} \right) \frac{d(\Delta E/E)^2_{tot}}{dt} \right\rangle - 2\alpha_x \varepsilon_x,$$

with similar equation for vertical motion, where defined

$$H_{x,y} = \frac{1}{\beta_{x,y}} \left( \eta^2_{x,y} + (\beta_{x,y}\eta'_{x,y} - \frac{1}{2}\beta'_{x,y}\eta_{x,y})^2 \right),$$

Partial decrements $\alpha_{x,y,s}$ defined as usual $\alpha_i = \frac{J_i}{2\tau_s}$, where $J_x \cong 1, J_y = 1, J_s \cong 2, J_x + J_s = 3$.

The energy spread is

$$\frac{d(\Delta E/E)^2_{tot}}{dt} = \frac{d(\Delta E/E)^2_{IBS}}{dt} + \frac{d(\Delta E/E)^2_{QE}}{dt} - \alpha_s \left( \frac{\Delta E}{E} \right)^2,$$

where energy spread from SR is

$$\frac{d(\Delta E/E)^2_{QE}}{dt} \cong \frac{55}{48\sqrt{3}} \frac{cr_0^2 \cdot \gamma^5}{\alpha |\rho|^3}.$$

In simplest model, IBS energy spread heating source could be described as

$$\frac{d(\Delta E/E)^2_{IBS}}{dt} \cong \frac{cNr_0^2 \ln_C}{\gamma^3 \varepsilon_x \sqrt{\varepsilon_y \beta_y} \sigma_b} \cdot \frac{1}{\sqrt{[1+\eta^2(s) \cdot (\frac{\Delta E}{E})^2 /(\varepsilon_x \beta_x)]}},$$

where $\ln_C = \ln(\rho_{max}/\rho_{min}) = \ln\sqrt{(v'/c)^6 / 4\pi \cdot r_0^3 n'_e}$ – is Coulomb's logarithm, $v'$ –is a transverse velocity, $n'_e$ –is density of particles in it's rest frame, $\sigma_b$ – is the bunch length, $\eta(s)$ is dispersion function here. Existing computer codes use more detailed representation of Coulomb's logarithm.

Let us find the beam parameters for different wigglers pattern. We start first with nine wigglers installed periodically around the model of the ring in straight sections close to the defocusing quads. The ring model has only regular part of CESR without hard bends. Number of 6.575 *meter* long dipole magnets in model is 72 with magnetic field ~0.892 *kG* @ 2*GeV*. Number of quads is also 72.
Wigglers in a model are tapered with magnetic field amplitude +¼,-¾,+1,-1,+1,-¾,+¼ .

Latest version of the wiggler tapering has values: +½,-1,+1,-1,+1,-1,+1,-1,+1,-1,+½

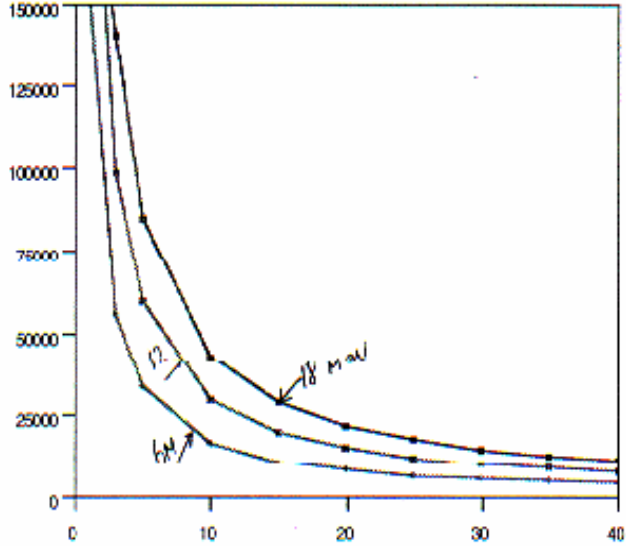

Beam lifetime, *sec* as a function of current in individual bunch, *mA*. Beam coupling $\varepsilon_z/\varepsilon_x = 0.1$. Three curves represent three different RF voltages: 6, 12 and 18MeV respectively.

Nine 3.6 T wigglers generate energy spread $\sim \Delta E/E \cong 1.16 \cdot 10^{-3}$. Emittance goes to $\varepsilon_x \cong 3.9 \cdot 10^{-6} cm \cdot rad$ @ $3mA$ per bunch. $\sigma_s \cong 1.7 cm$ @ $6 MeV$, $\sigma_s \cong 1.23 cm$ @ $12 MeV$ and $\sigma_s \cong 1.cm$ @ $18 MeV$ of RF field. Damping time ~26 ms.

In the Table below we summarized some parameters as a function of the number of wigglers. IBS included in each emittance for $3 mA/bunch$.

| Number of wigglers | $B_{max}$, T= | $\Delta E/E \times 10^4 =$ | $\varepsilon_x \times 10^6$, $cm \cdot rad =$, |
|---|---|---|---|
| 9 | 3.6 | 11.6 | 3.9 |
| 18 | 2.4 | 9.4 | 9.7 |
| 36 | 1.6 | 7.7 | 5.5 |

Relative emittance increase at 18-wiggler structure can be explained by increase of envelope functions, as there was no adjustment of structure made.
For final answer to the question about emittance, it is necessary to insert wigglers in realistic CESR model with real free spaces available.
In a model the wiggler field was represented as a step like functions. In reality the longitudinal field distribution is a sin-like function so effective damping for the maximal field indicated will be less.

## FOCUSING BY RF CAVITY

Let us make some comment here, which might be important for successful operation of CESR at low energy. We will approximate the CESR cavity by a pillbox one.
The pillbox cavity has the fields as the following
$$E_s \cong E_0 J_0(r/\lambda_{RF}) Cos\omega t$$
$$H_\varphi \cong i\frac{E_0}{c} J_1(r/\lambda_{RF}) Sin\omega t.$$

The transverse momenta, obtained by the particle goes to
$$\Delta p \cong ev \times BL/c \cong eBL,$$

where $L$ is effective length of cavity. So the angle obtained goes to

$$\alpha \cong \frac{r}{F} \cong \frac{\Delta p}{p} = \frac{eE_0 L}{2pc} \frac{r}{\lambdabar} Sin\omega t,$$

so the focal distance of RF field $F$ goes to

$$\frac{1}{F} \cong \frac{eE_0 L}{2pc} \frac{1}{\lambdabar} Sin\varphi.$$

For $\lambdabar \cong 10 cm$, $eE_0 L \cong 20 MeV$,

$$F_{RF} \cong \frac{2 \cdot 2000}{20} 10 \cong 20 m.$$

For the reference, the quadrupole lens with $G \cong 0.5 kG/cm$, a lens of $l=60cm$, for the 2 GeV beam, having magnet rigidity $HR \cong 6 \times 10^3 kG \cdot cm$

$$F_{MARK} \cong \frac{HR}{Gl} \cong 200 m.$$

This question needs to be investigated more carefully.

**FREE SPACE INVENTORY**

Free space (flange to flange) inventory of CESR chamber is the following:

| **East:** | | **West:** | |
|---|---|---|---|
| E2-Q8 | ~ 100 cm | Q9-RF | ~ 150 cm |
| Q9-… | ~ 400 cm | Q9- wall | ~ 200 cm |
| Q14… | ~ 200 cm | Q14 … | ~ 200 cm |
| Q18-Q19 | ~ 460 cm | Q19-Q18 | ~ 460 cm |
| Q29… | ~ 100 cm; skew Q here | Q36… | ~ 100 cm |
| Q46-dip | ~ 100 cm; pickup here | Q43-Q43A | ~ 180 cm |
| Q45-sep | ~ 120 cm; Oct, here | Q45… | ~ 150 cm; Oct. here |
| Q43-Q43A | ~ 180 cm | Q46… | ~ 100 cm; CERN monitor here |

Total =  1660 cm=16.6 meters          Total = 13400 cm=13.4 meters
    + Wiggler~280 cm                   +Wiggler~280 cm

Total ~30 meters + Two wiggler places (5.6 m)

This space allows installation of many independent ~ 1 meter long wigglers.
Initially the number of wigglers might be 5+5 (or five at each East/West side of CESR ring).

# WIGGLER PROTOTYPE

**3D** calculations of the wiggler prototypes done. Number of poles (North and South pairs) in models was seven, six, nine and eleven respectively.

**Seven pole** wiggler, has poles displaced with period 28 *cm*. Total length goes respectively to $14 \times 7 = 98$ cm. Each pole has width in transverse direction as 24 cm. For each coil reserved thickness of 1 cm with height about ~3.9*cm*. The pole gap: $\pm 34.78$ *mm* = 69.57 *mm* total opening possibility for installation of warm vacuum chamber inside the wiggler with height open to the beam up to ~55mm. Number of turns with existing wire with diameter $D = 0.017'' \equiv 0.4318mm$, is about 1889 with filling coefficient ~0.7.

The wiggler is tapered for zero orbit displacement with current in the coils 39.2; 180; 240 *kA/pole*; The current in a wire in central pole is about 130*A*. This current gives peak field about 3.6*T* in the middle of the central gap and about 6 T in the coil in places with maximum field strength. Meanwhile critical current for 3T is 240 A; for 5 T field the critical current is 165 *A*; for 7 T the critical current is 100*A* for this wire. So this is much above required, as in real situation the wiggler will work at ~2T peak field.

HTS leads will be used to deliver the current to the coil.

Suggested that end poles will have additional trimming coils, or even independent power supply. As the currents and fields are far from critical, the problems with quenches will be minimized. Quench protection supposed to be done with usual way: by active and passive systems.

Operation with captured flux could be recommended also. This will drastically reduce the heat losses and requirements for the power supply.

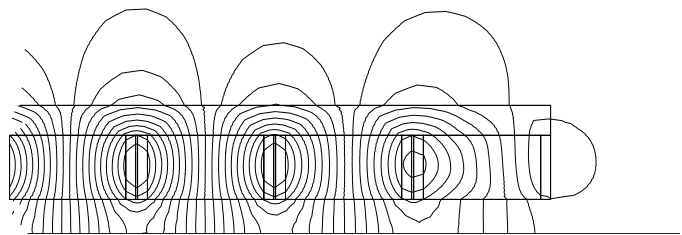

Map of lines close to the end part

Field map around central pole digitized. Absolute values of the field are shown.

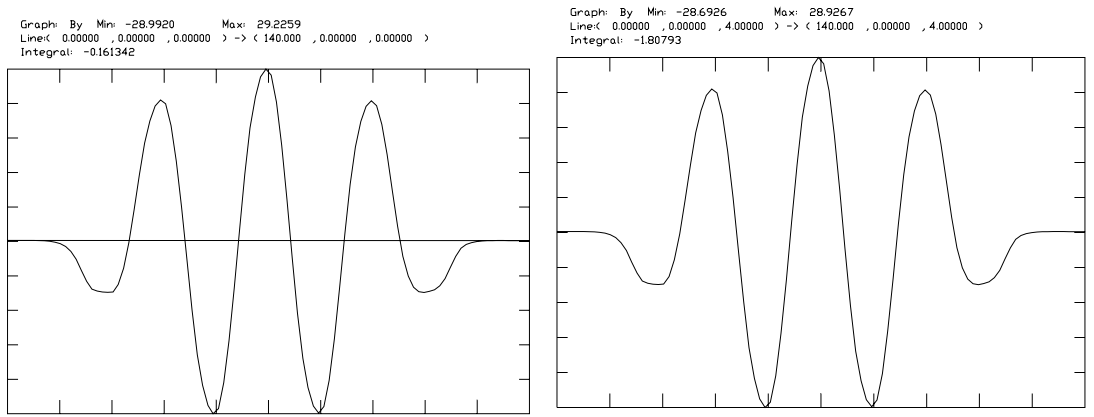

Left: Longitudinal field distribution along central line.  Right:   Longitudinal field distribution +4cm off central line.

Transverse distribution shows, that the field variation at the distance ±48 *mm* off axis, has the field lower by ~2%, and, hence, intensity variation ~4% for the particles, running through the center of the wiggler and at the edge of aperture.

The integral of magnetic field along the wiggler is a function of transverse position of trajectory. This dependence could be represented as the following

$$\int_{-\infty}^{+\infty} B_y(s)ds \cong -0.146 - 0.087 \cdot x^2 - 0.00134 \cdot x^4 + 1.683 \cdot 10^{-5} x^6 + ... \ .$$

The first term: −0.146 could be eliminated by adjustment of currents in edge poles however. For the reference, CESR sextupole at 1A feeding current gives $\int_{-\infty}^{+\infty} B_y(s)ds \cong 0.1065 \cdot x^2$. Polarity of this sextupole term could be adjusted to the necessary one by reversing of the field in coils, as the damping rate does not depend on the field polarity.

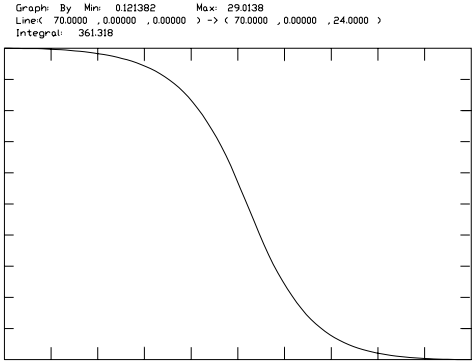

Transverse field distribution across the central pole. Horizontal axis runs from the middle of wiggler to 24cm outside the central line. Particles are running perpendicular to the plane of the figure.  The width of the pole is 24cm: ±12 cm from the central line, correspond to the center of coordinates on the drawing.

For **six pole wiggler** the situation with transverse dependence of wiggler integral is better. The tapering at one end is the same as for seven pole wiggler, but at other end one full current pole is removed and replaced by tapering poles with reversed (with respect to first end) polarities of the field.

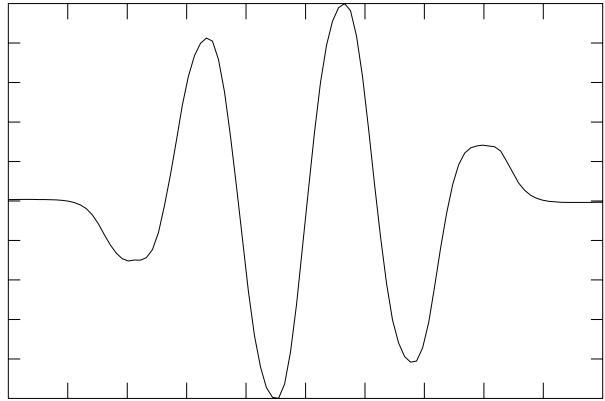
Longitudinal field distribution for six-pole wiggler. ~2.8 T in maximum.

The six-pole wiggler demonstrates transverse dependence like

$$\int_{-\infty}^{+\infty} B_y(s)ds \cong -0.06 - 0.034 \cdot x^2 + 0.00234 \cdot x^4 - 5.28 \cdot 10^{-5} x^6 + ... \ .$$

Tuning of this wiggler for zero transverse displacement will be much more difficult, than odd-pole wiggler, however. This does not eliminate the dependence of power losses on transverse position in a wiggler.

Next prototype model will have reduced length of the pole in edge poles. This will reduce the total length of the wiggler and will make the field in these poles less sensitive to the iron poles saturation. The height and width of the coils will be optimized also. Iron will be added to the yoke.

**Nine pole wiggler** field distribution is represented in the Fig below. Total current in the central pole coils is 240 *kA* in total cross section area ~3.6$cm^2$. This current gives the field strength maximum ~2.9*T* at the axis and ~6*T* in places with maximum field, located inside the coil. The gap between poles is 70 *mm*, pole width 240 *mm*, period 200 *mm*.

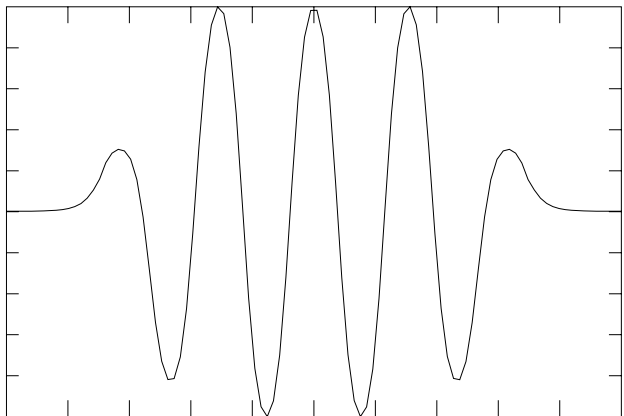
Longitudinal field distribution for 9 pole wiggler. Wiggler has period of 200 mm and pole width 240 mm.

All parameters (damping time, focusing) do not depend on wiggler period at all. The lengthening of trajectory is $\Delta \cong \pi K^2 \lambdabar_w / 2\gamma^2$ is going in favor of lowering period. The opening angle of radiation is $\sim K/\gamma$, what helps in avoiding overheating the inner wall of regular vacuum chamber. $K$ factor for these parameters are $K = eH_w \lambdabar_w / mc^2 \cong 93.4 \cdot H_w[T] \cdot 2\pi \lambdabar_w[m] = 93.4 \cdot 2.9 \cdot 0.2 = 54$ (instead of 75.6 for 28 $cm$ period of seven pole wiggler) for the same magnetic field.

**Eleven-pole wiggler** is preferable candidate for final choice. Some parameters of this wiggler are represented in Figs below.

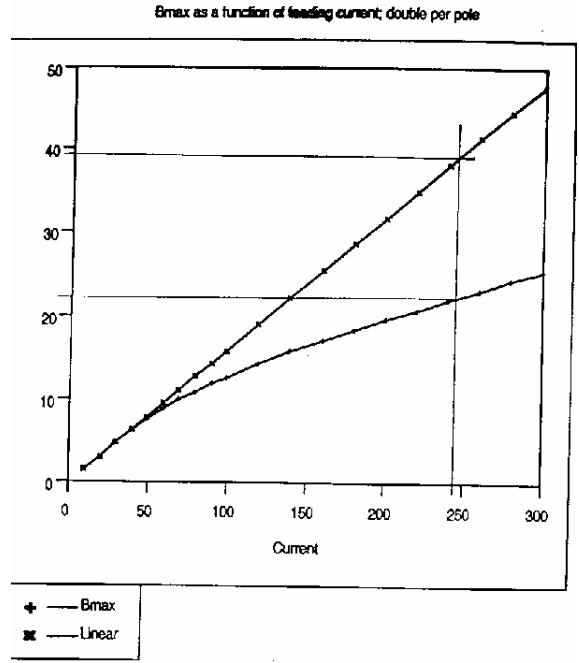

Field in a gap as a function of current

The figure above shows the saturation curve behavior. So at working point the field is roughly a half of its possible value, if the iron is not saturated.

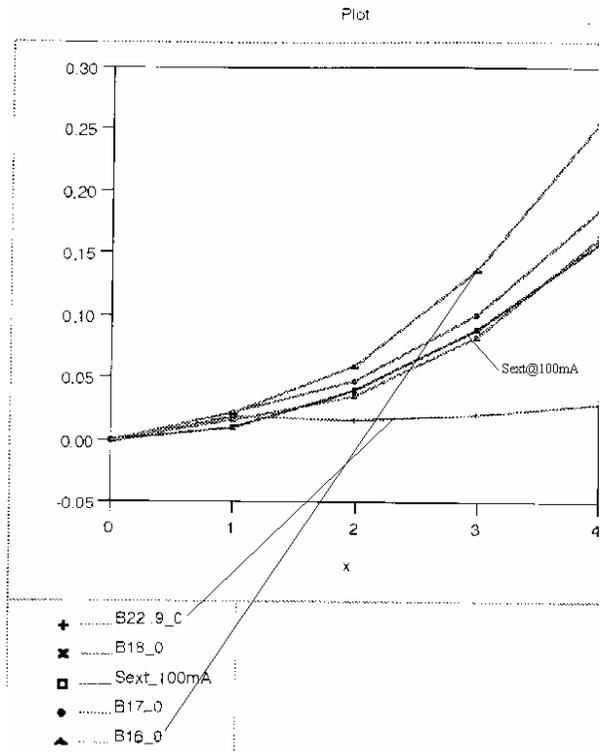

Calculated transverse dependence of integral, *kG·cm* along longitudinal coordinate for 11 pole wiggler as a function of current. Also shown integral of CESR's sextupole running at 0.1A.

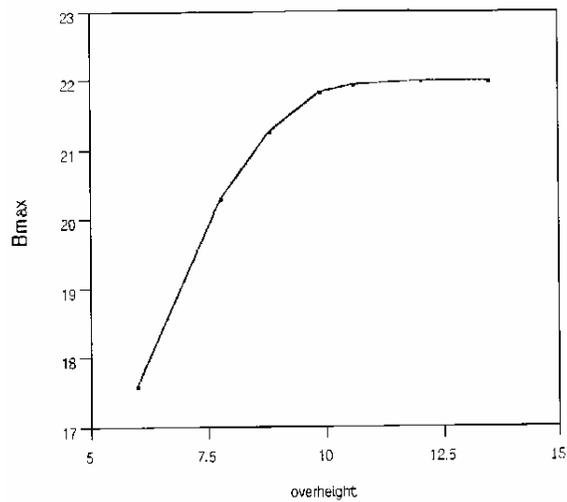

Field in the gap as a function of iron thickness at the top of wiggler yoke.

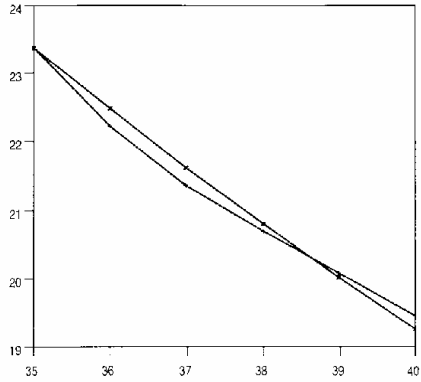
Field in the gap as function of the gap for fixed Amper/turns (120kA/turns).

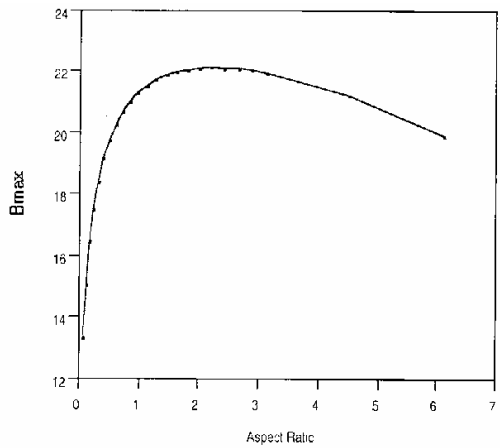
Field in the gap as a function of the coil aspect ratio

Coils aspect ratio is $0.75'' \times 0.75''$ (shifted out of optimum). This is done for better winding.

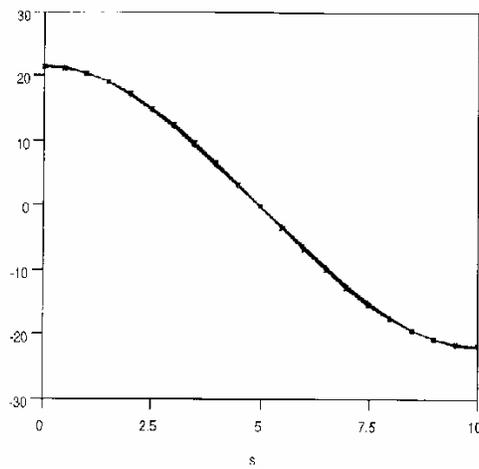
Field as a function of longitudinal coordinate. Sine curve in drawn for reference also.

Stored energy in 11 pole wiggler is 70 kJ at the field amplitude ~22kG.

|                       | # turns/coil | $L_{wire}$/coil | Current, A |
|-----------------------|--------------|-----------------|------------|
| Wire dia 0.7mm        | 506          | 346             | 247        |
| Wire dia 0.6mm        | 628          | 444             | 192        |
| Wire dia 0.381mm(SSC) | 1558         | 1089            | 80         |
|                       |              |                 |            |

The current indicated is half of critical for the field indicated. The wire 0.6 mm will be used in Dewar 3 pole wiggler also. Wire 0.7 mm suggested to be the one for the working prototype.

## HEAT LOSSES

The wiggler will have room temperature inner vacuum chamber made on copper, having the same shape as CESR' s one. This inner chamber has water jacket for SR absorption. This room temperature vacuum chamber and helium vessel will be interlaced by shield cold by Nitrogen or by Helium vapors.

Synchrotron radiation from the wiggler itself does not touch the wall of inner chamber. Cooled jacked will absorb SR from neighboring wigglers installed in series in long straight sections and from the magnets.

Utilization of cryo-coolers could be recommended after consideration of the transfer line cost. This might be important only for the wigglers installed closer to the North area. Typical cryocooler (RDK-415) delivers 35 *W* at 50 *K* in first stage and 1.5 *W* at 4.2 *K* at second stage at 50 *Hz* operation frequency. At 60 *Hz* corresponding figures are 45 and 1.5 *W*, respectively. Maintenance interval ~10*khours*. The cryocooler could be removed with minimal losses of helium, if necessary.

The heat losses defined by losses on support system, radiation, current feadthrough and heat transfer by residual gases.

Heat transfer by support system defined by

$$Q = -\frac{\int_{T1}^{T2} k(T)dT}{\int_{x1}^{x2} \frac{dx}{A(x)}},$$

where $k(T)$ is thermal conductivity coefficient depending on temperature, $T_{1,2}$ is end temperature respectively, $A(x)$ is an area of supporting element as a function of distance along the support. Integral $\int_{T1}^{T2} k(T)dT$ is well tabulated for numerous materials. For support (or suspend) of constant cross section the last formula stands

$$Q = -\frac{A}{l} \int_{T1}^{T2} k(T)dT,$$

where l is linear dimension of supporting/suspension element. For example for glass-type material $\int_{4.2}^{77} k(T)dT = 0.175 W/cm$ and for idealized element having the shape of cube by $1 \times 1 \times 1\,cm^3$, made on glass –type material (G-10) the heal losses comes to $0.175W$. It is easy to normalize this number to any real size of support.

For example, if support has a shape of a thin-wall cylinder of ~6 *cm* outer diameter with 3*mm* wall thickness, the cross section will have area ~$6cm^2$. If this support is about 6 *cm* long and total number of supports is three, then total heal flow comes to $3 \times 0.175W = 0.525W$ total.

Not indicated here the losses associated with filling tubes, control wiring (level, temperature…) and losses for the trimming coil leads. As the last ones supposed to work at small current (~mA level), the losses could made below 40 *mW/pair*.

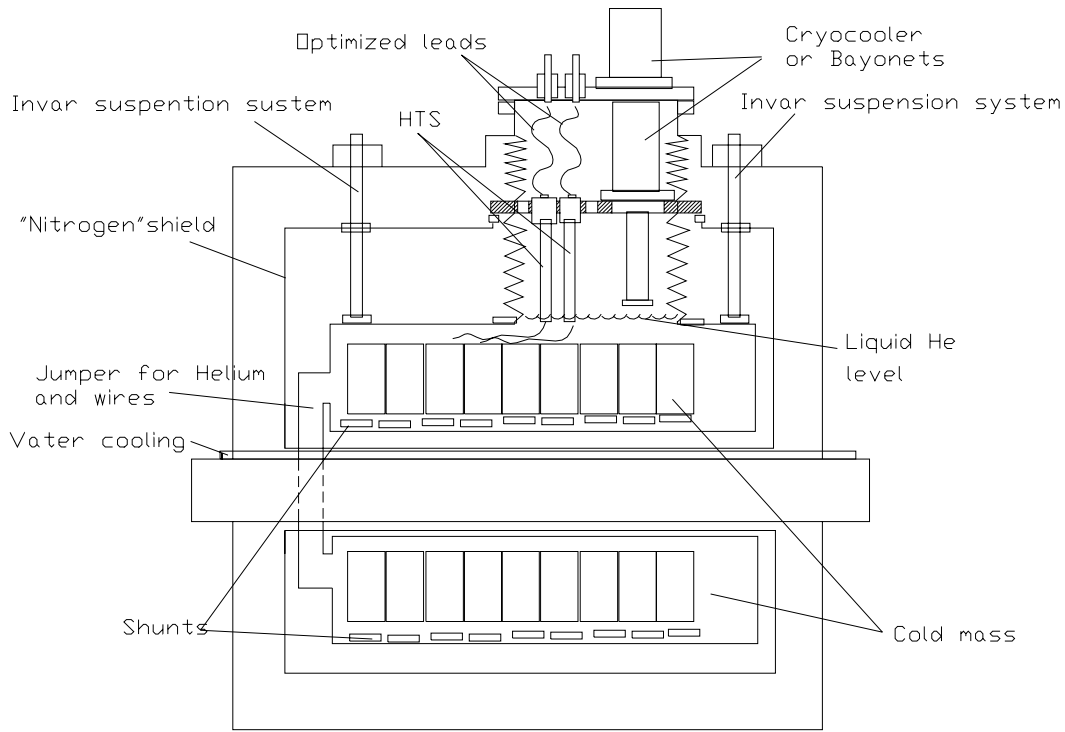

Concept of cryostat with suspension

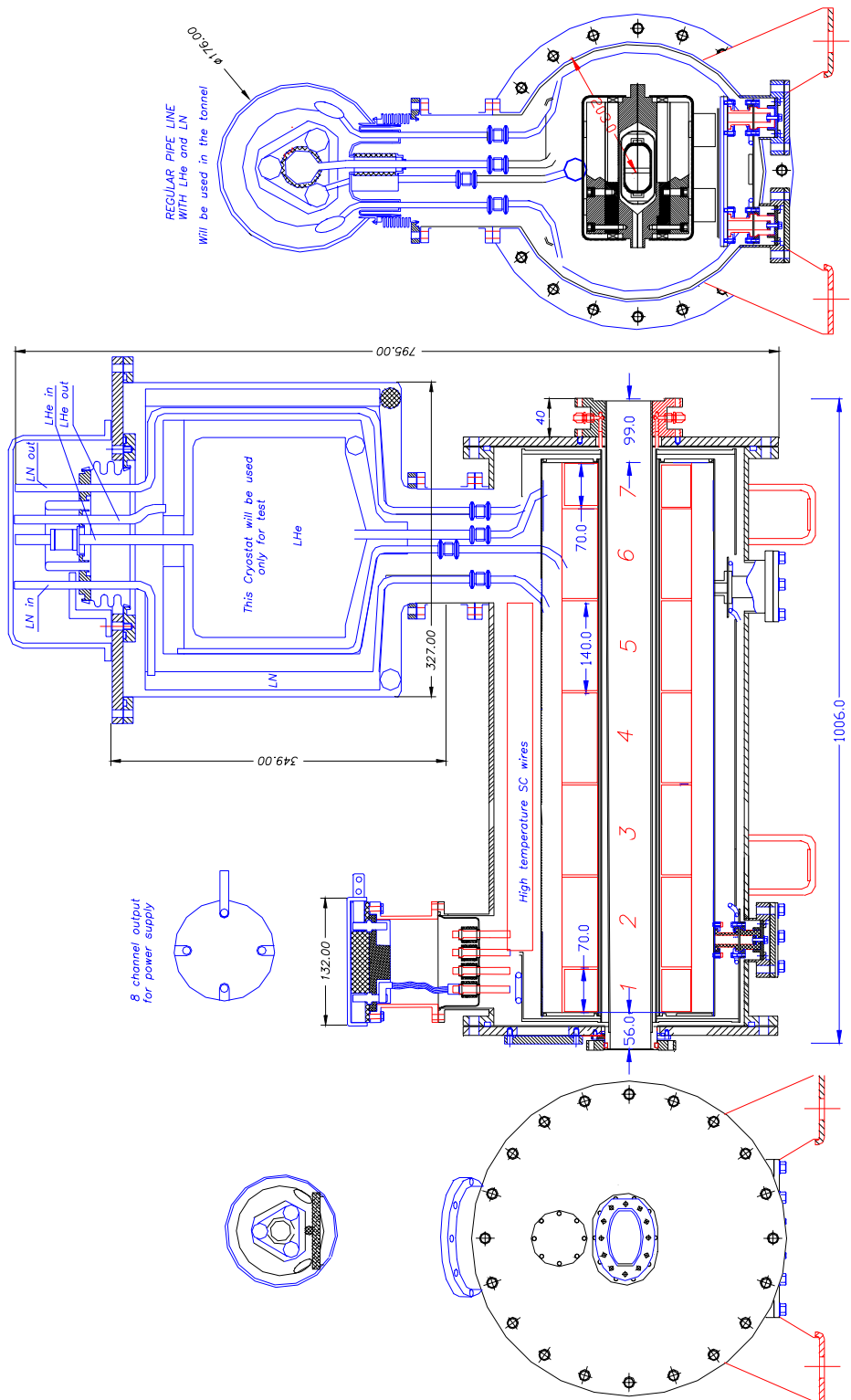

Wiggler Cross-section. Seven pole wiggler is shown here.

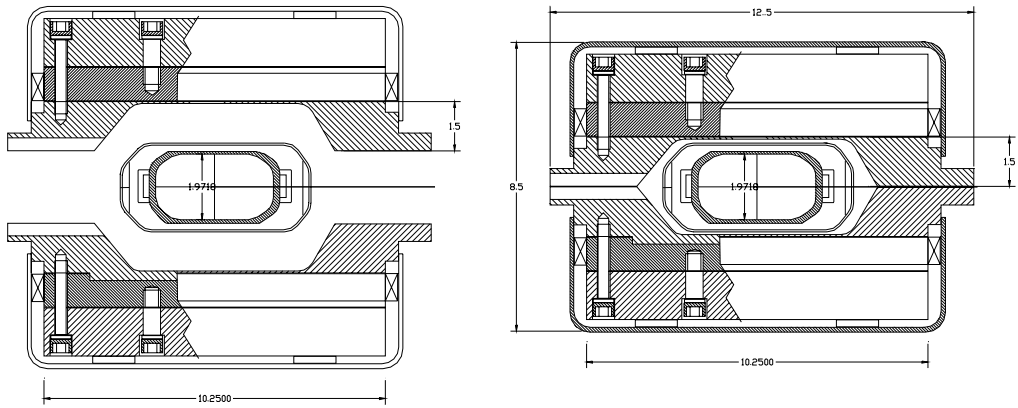

Cold mass schematics.

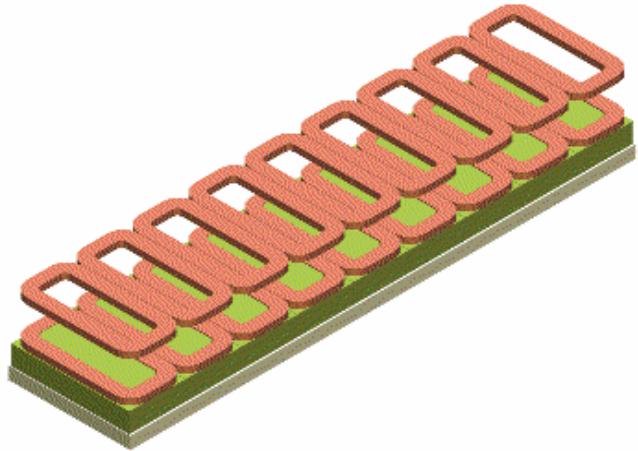

Coils and lower pole array

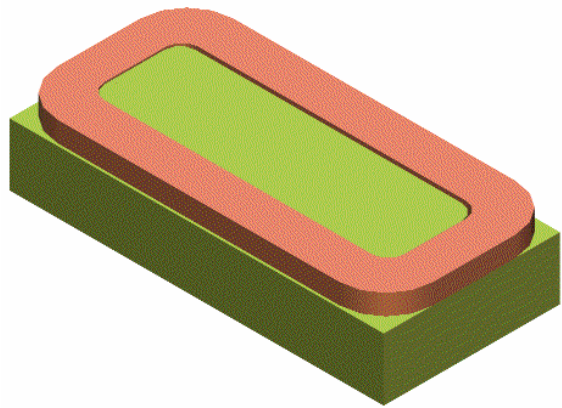

The pole module. Coil and yoke.

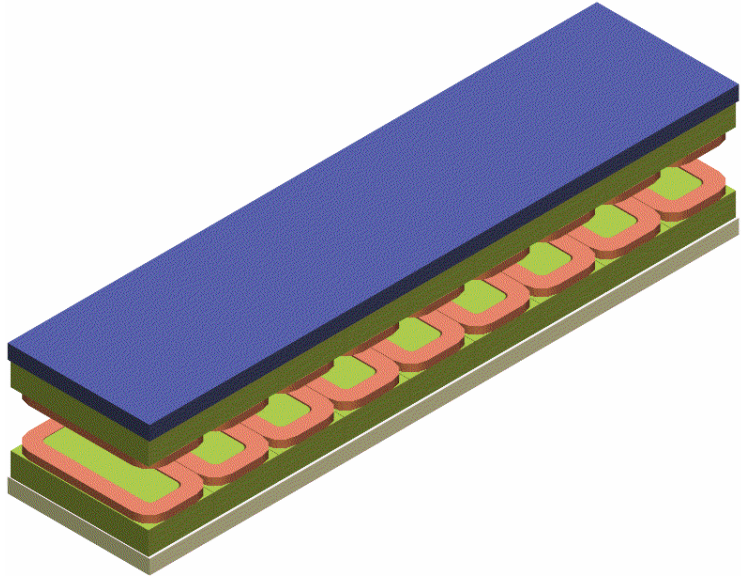

Poles.

For calculation of losses by radiation we have the formula, valid if the surfaces have about the same area, as

$$P[W/m^2] = \frac{5.67 \times 10^{-8}}{1/\varepsilon_1 + 1/\varepsilon_2 - 1} \times (T_1^4 - T_2^4),$$

where $\varepsilon_{1,2}$ and $T_{1,2}$ are corresponding emissitivies and temperatures of each surface respectively. Temperatures under interest are 4.2K, 77.6K and 300K. With last formulawe have $P_{He-N2} \cong \varepsilon_{eff} \times 2[W/m^2]$ and $P_{N2-300} \cong \varepsilon_{eff} \times 457[W/m^2]$. Inner area around central tubing can be estimated as ~0.3 $m^2$ and respectively outer area around cold mass of magnet as 1.2 $m^2$. So $P_{He-N2} \cong \varepsilon_{eff} \times 0.6[W]$ and $P_{N2-300} \cong \varepsilon_{eff} \times 137[W]$ for inner tubing and $P_{He-N2} \cong \varepsilon_{eff} \times 2.4[W], P_{N2-300} \cong \varepsilon_{eff} \times 550[W]$ for outer side around cold mass.

Emissitivity for polished copper, for example, is around $\varepsilon \cong 0.023$, and about the same for aluminum. For polished StSteel this is $\varepsilon \cong 0.074$. Polished walls delivering moderate emissitivity $\varepsilon \cong 0.1$ together with ~10 layers of superinsulation in inner tubing yield $P_{He-N2} \cong 0.6 \times 0.1 \times \frac{1}{9} \cong 6.7 \times 10^{-3}[W]$, $P_{N2-300} \cong 137 \times 0.1 \times \frac{1}{9} \cong 1.5[W]$. For outer side we can suggest 20 layers of superinsulation what delivers $P_{He-N2} \cong 2.4 \times 0.1 \times \frac{1}{20} \cong 0.012[W]$ and $P_{N2-300} \cong 550 \times 0.1 \times \frac{1}{20} \cong 2.8[W]$. So total radiation losses for helium comes to

$$P_{He} = P_{inner} + P_{outer} + P_{current\ leads} \cong 6.7 \times 10^{-3} + 0.012 + 0.18 \equiv 0.2W,$$

where we included the losses associated with current leads what is 180 $mW$ per pair of 25 cm long 200 $A$ current leads of ASC. Together with losses on support system it goes to

$$P_{He,tot} \cong 0.53 + 0.2 = 0.73W.$$

Together with uncalculated ones this goes to ~1$W$ of total Helium heat losses per cryostat. This figure within reasonable cost margins for 1 or 1.5 W cryocooler system in a favor of 1.5$W$ as the cost will be only 8k$/wiggler higher, but allows more losses.

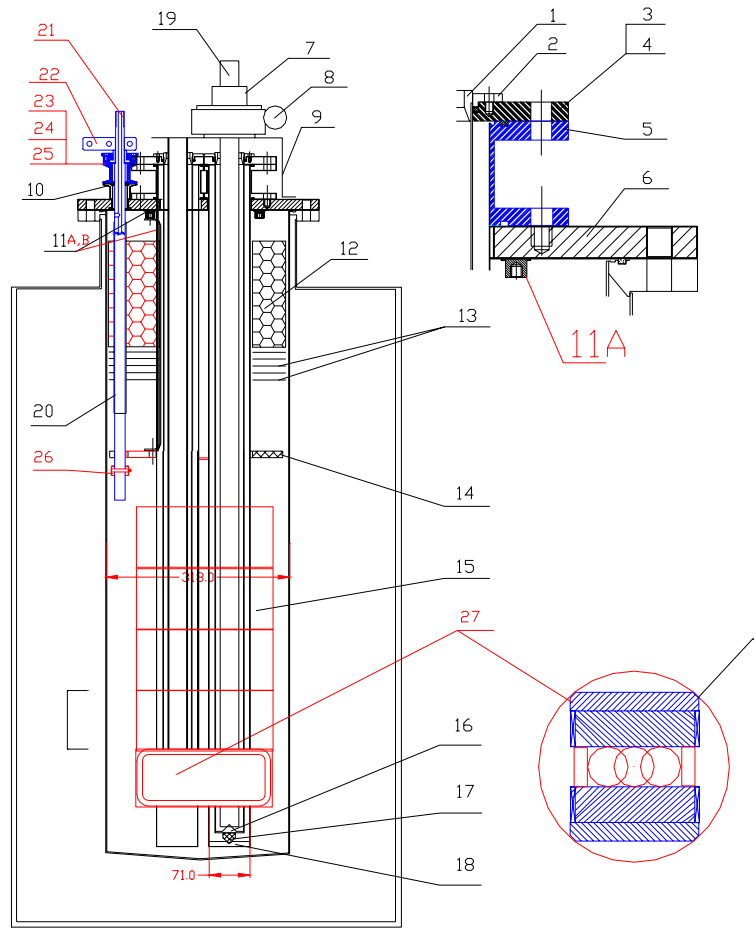

Model of wiggler prepared for testing in existing cryostat. Up to five poles can be inserted in existing Dewar.

The wiggler coils could be tested in liquid helium within existing cryostat. Size of this cryostat allows testing up to four-pole wiggler.

To cool down a ~600kg $\cong 1500 lb$ of cold mass, it will be required $V \cong 0.36[L/lb] \times 1500[lb] \cong 540 Liters$ of liquid Helium, if enthalpy of gas used. To reduce this amount, preliminary cooling by Nitrogen can be used. If this cold mass cooled preliminary by Nitrogen, the amount of liquid Helium goes to $V \cong 0.07 \times 1500 \cong 105 Liters$.

## SUPPORT/SUSPENSION SYSTEM

Support system must hold the cold mass in proper position, when cooled. The wight of the cold mass estimated as ~560 kg of steel and ~28 kg of copper wire. So the mass around 600 kg can be taken for estimation. This is ~200 kg/per leg.

The problem here is a large temperature diapason what requires a proper consideration of expansion. This is important as the thermal shrink reaches $\Delta l \cong 11 \times 10^{-6} \times 1000 \times 300 \cong 3 mm$ for iron, and $\Delta l \cong 17 \times 10^{-6} \times 1000 \times 300 \cong 5.1 mm$ for StSteel container. Supporting system is a three and half leg system, proposed for dual aperture magnet, which allows stabilization of the wiggler magnet motion during the cooling. Principle of support system is represented in figure below.

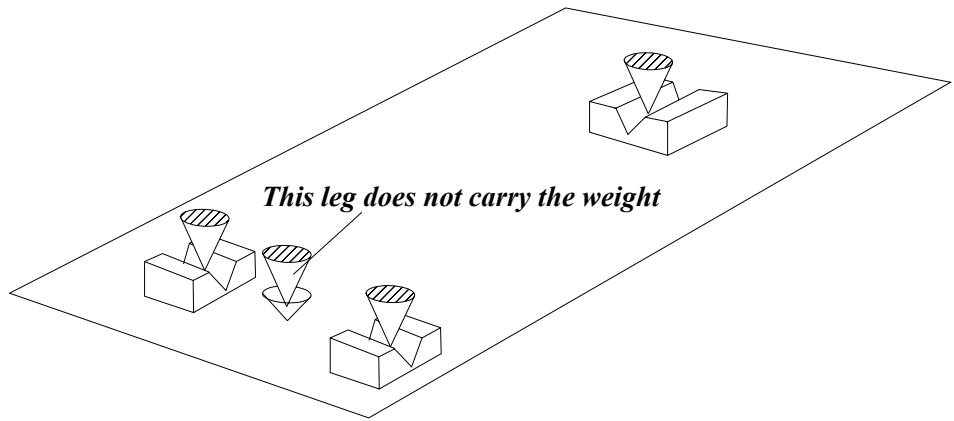

Principle of stabilization of cold mass position during the cooling.

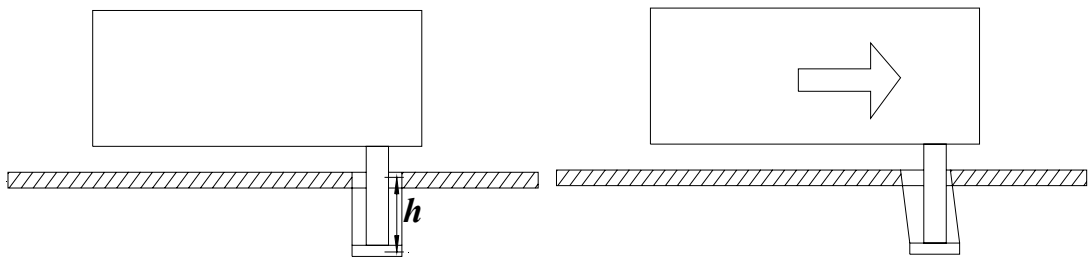

Swinging support. Technical realization of the moving leg from previous figure.

Variation of the height is going here is second order if compared with the lengthening. So if the lengthening is $\Delta$, the height variation is $\Delta h \cong \dfrac{\Delta^2}{2h}$, where $h$ is the height of swing. The lengthening after cooling cold mass *and* intermediate support down to the temperature ~70K, gives $\Delta$~1.16*mm*, *h*~60*mm*, $\Delta h \cong \dfrac{1.2}{2 \cdot 60} \cong 0.01 mm \cong 10 \mu m$. This figure is acceptable. In principle material of the swinging strips can be chosen to compensate the height variation. Design allows to adjust the vertical position of the cold mass during the tuning the wiggler, *in situ*.

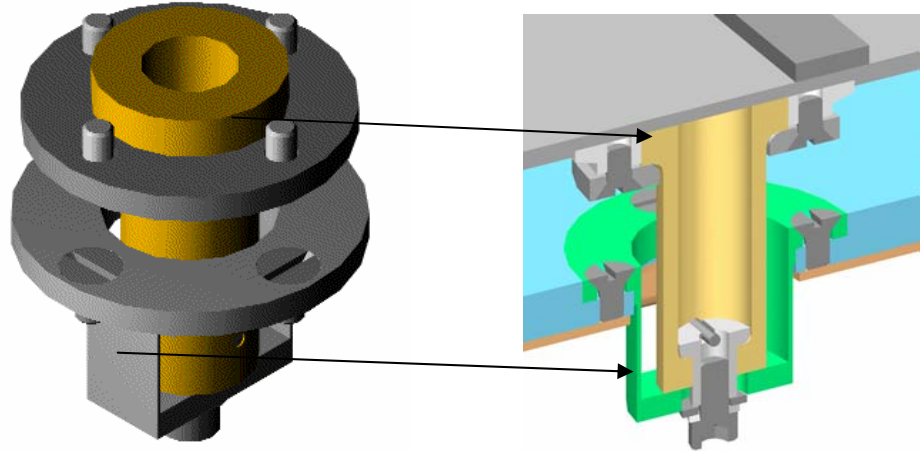

3D view of the leg assembly.

The cylinder has ~6 cm in diameter, thickness~3mm and the length ~6cm. The strips made on stainless steel add the resistance to the heat loss. So the estimation of power losses ~0.175 watts/leg made without these strips, guaranteed by the strips here. The cylinder itself is strong enough to hold 200 kg/leg weight.

Suspension system is less likely here, as the cryocooler can initiate vibration of the cold mass.

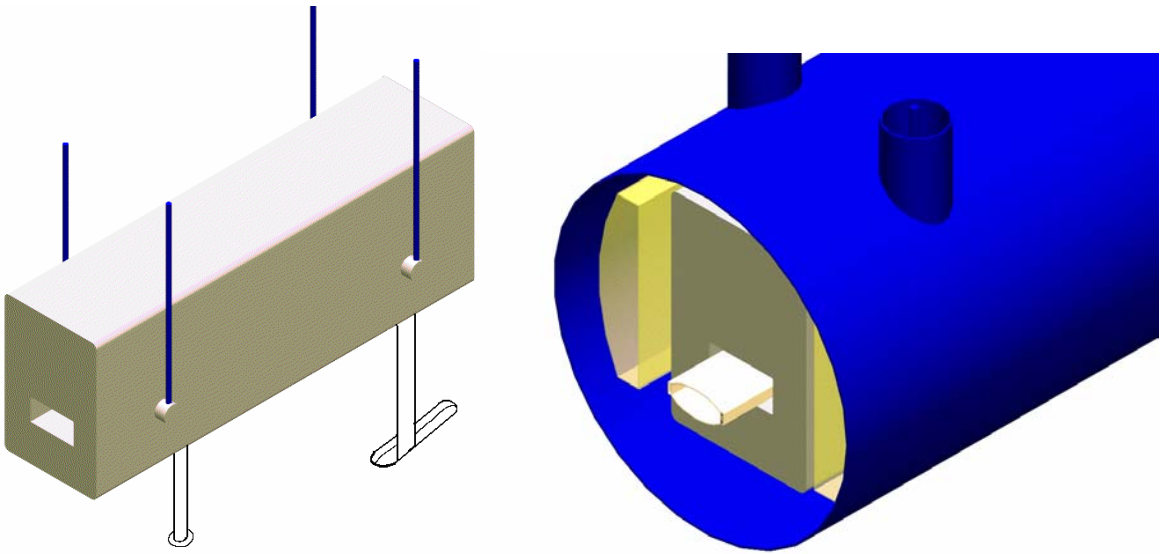

Principle of cold mass stabilization in case of suspension (left). The design allows adjustments of cold mass position in vacuumed cryostat.

## WIGGLER IN A TUNNEL

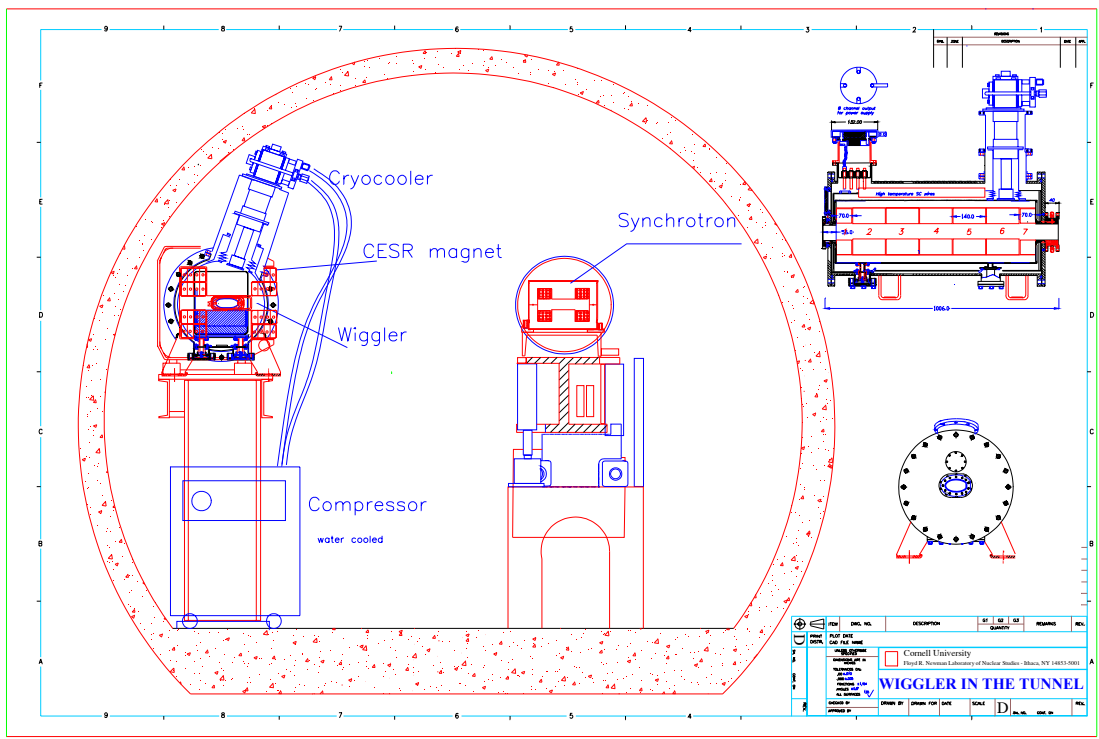

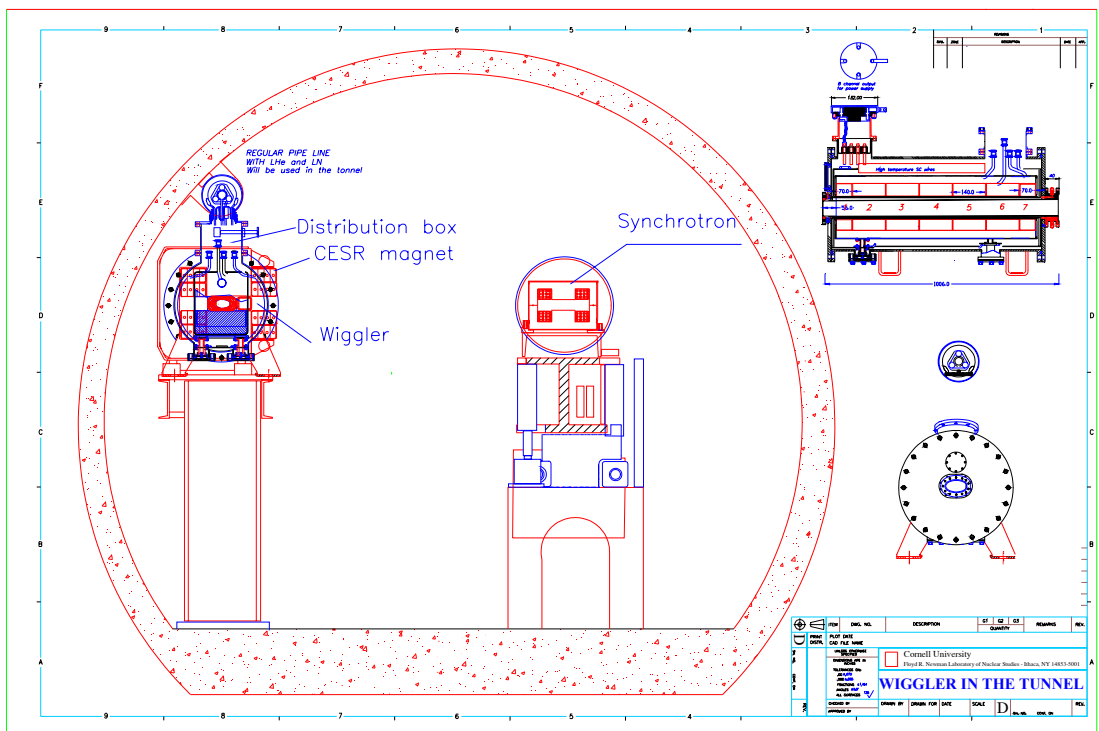

The wiggler positioning in the tunnel.  Upper figure shows the cryocooler and compressor. Lower figure shows the cooling method with cryoline running around the tunnel.

We can expect both- cryoline and cryocooler type cooling for the wiggler.

# QUENCH PROTECTION ELEMENTS

For quench protection few systems will be used.

**Passive cold shunts**.
Each coil will be shunted by copper/StSteel shunt.
For the ratio of Cu/NbTi 1.35/1, the area occupied by wire $S = 3.8 \times 10^{-3} cm^2$ will be distributed between Copper: $S_{Cu} = 2.18 \times 10^{-3} cm^2$ and NbTi $S_{NbTi} = 1.62 \times 10^{-3} cm^2$.
Resistance of the coil will be

$$\frac{1}{R} = \frac{1}{R_{Cu}} + \frac{1}{R_{NbTi}}$$

As the length of the coil wire will be $N \sim 346$ meters, the resistance of Copper part will be at room temperature

$$R_{Cu} = \frac{\rho_{Cu} \times L_C}{S_{Cu}} \cong \frac{1.7 \times 10^{-6} \cdot 34600}{2.18 \times 10^{-3}} \cong 27\Omega.$$

Resistance of NbTi part (at ~ room temperature) will be

$$R_{Cu} = \frac{\rho_{NbTi} \times L_C}{S_{NbTi}} \cong \frac{13.3 \times 10^{-6} \cdot 34600}{1.62 \times 10^{-3}} \cong 284\Omega,$$

so the effective resistance will be ~24.6 Ω. At lower temperature this value will be less, of cause. So the shunt, say 0.12 Ω will recapture ~99.5% of current at room temperature, and the current in the coil will be 1.2 A only.

**Fast switch** represented in Fig. below.

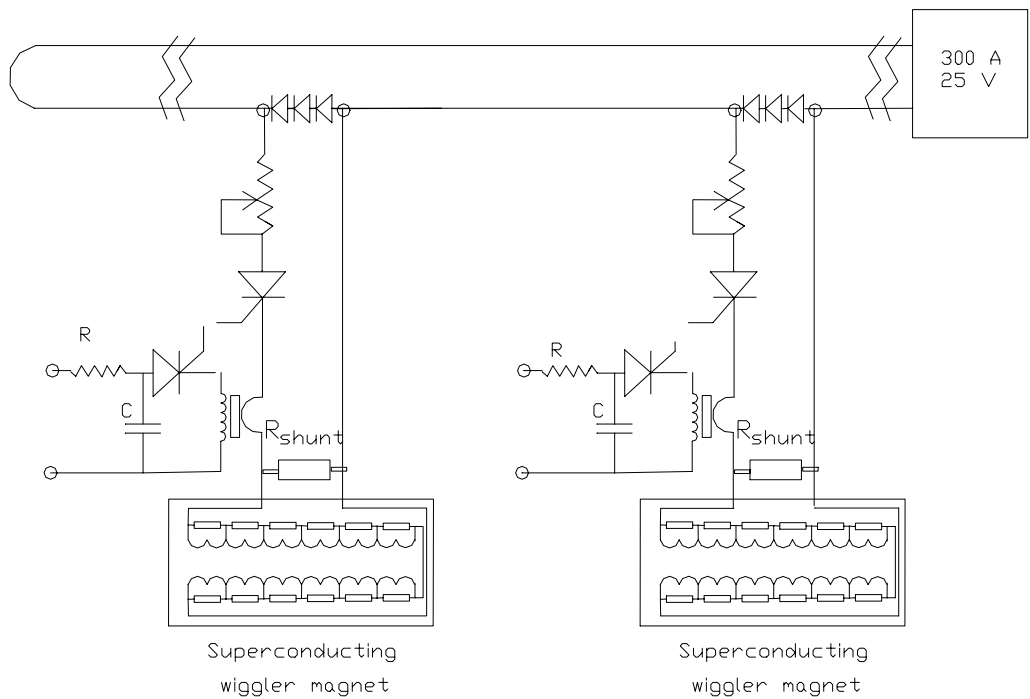

**FIGURE 1:** Fast switch for turning off the magnet and power supply principle.

Transformer introduces the current of opposite direction into the circuit. So secondary windings are made able to carry a DC current 250 A. This system is compact and faster, than mechanical one. Voltage developed across $R_{shunt}$ is $V = I_{wiggler} \times R_{shunt}$.

**Quench diagnosis** system for fast off.

A wire low inductive loop will be used for monitoring of developing quenches.

## COST ESTIMATION

At the moment we have the following figures.

**BINP,** *Novosibirsk*, is ready to manufacture a first prototype for 450k$. The same set for mass production of 20 units estimated as 250k$ per magnet.

This price includes:
- magnet
- cryostat
- cooling machines (first stage -41 Watt at 50K, second stage 1 Watt at 4.2 K; and other one for shield cooling - two stage cooler: first stage -80 Watt at 70 K, second stage- 14 Watt at 20 K). If just cooler for shield cooling the wiggler is used, the price could be decrease down ~50k$)
- liquid helium filling system, quench protection system and vacuum chamber of the wiggler with photon absorbers.

Novosibirsk team has experience in manufacturing of similar products. It is possible to fabricate the wiggler without any coolers but using liquid nitrogen for shield cooling and price will decrease for 10-15k$. (BINP fabricated two wigglers like that: 7.5 T wiggler for Pohang Light Source and 7 T wiggler for Louiziana State Universisty, CAMD). Recommendation is do not use liquid nitrogen for shield cooling; coolers are much more convenient and reliable enough.

**Advanced Design Consulting**, **ADC**, local company, is ready to *design* the wiggler for ~25k$. Prototype will cost~150k$ and will include
- central chamber, copper, water cooled
- nitrogen chamber, copper
- helium chamber, copper
- helium magnet chamber, stainless steel
- pole iron
- magnet winding
- nitrogen magnet chamber, stainless steel
- high-temperature superconductor wires
- magnet support bars, stainless steel
- magnet support bar spacers, stainless steel
- power feedthrough
- cold finger feedthrough, nitrogen
- cold finger feedthrough, helium
- isolation supports
- feet
- clean weld of assembly
- outer chamber
- end flanges
- magnet wire
- DISPLEX cold finger, nitrogen

- DISPLEX cold finger, helium

Production cost:
| Qty. | $each | Total cost |
|---|---|---|
| 20 | $118,960.00 | $2,379,200.00 |

**JANIS** Corporatin.
 This company has experience in design of similar products. The cost of 1-m long cryostat designed by this company is about 110k$. This does not include the cost of cryo-coolers and the magnet itself. It is unlikely, that the cost of cryo-coolers will go down as result of order ~20 units. The cost of single cryocooler system -cold head and compressor- is ~44k$ per 1.5$W$ head for the water-cooled compressor from JANIS (36k$ for 1$W$ cooler).

## CESR MAGNETS

Field quality of CESR Mark-II quad magnets calculated on the basis of 3D codes indicates field quality (*s*- longitudinal coordinate)

$$\frac{\int_{-\infty}^{+\infty}[B_y(x,s)-G(s)\cdot x]ds}{x\cdot \int_{-\infty}^{+\infty}G(s)ds} \leq 6\cdot 10^{-4}, \ |x|\leq 4cm$$

for feeding current in range 10-50 *A*. This quality does not account the presence of neighboring sextupole.

## CONCLUSION

The strategy described with many short wigglers looks more attractive, if compared with long wiggler. Installed once, long (40*m*) wiggler will require significant non-reversible modification of CESR's hardware.

Short wiggler could be easily manufactured and tested. Serial production can be easily arranged on the basis of a single prototype test.

Preferable strategy in wiggler fabrication is a self-made prototype (with manufacturing some parts of cryostat in local companies). Further there will be significant freedom to make parts of cryostat elsewhere. Windings of coils and final assembly inside laboratory can safe significant amount of money.

200k$ per wiggler can be recommended as zero approximation.